\documentclass[twocolumn,showpacs,preprintnumbers]{revtex4}
\usepackage{amssymb}
\usepackage{amsfonts}
\usepackage{amsmath}
\usepackage{graphicx}
\usepackage{dcolumn}
\usepackage{bm}

\input{tcilatex}

\begin{document}

\title{Quasithermodynamic Representation of the Pauli Markov equation and
their possible applications}
\author{E. D. Vol}
\email{vol@ilt.kharkov.ua}
\affiliation{B. Verkin Institute for Low Temperature Physics and Engineering of the
National Academy of Sciences of Ukraine 47, Lenin Ave., Kharkov 61103,
Ukraine.}
\date{\today }

\begin{abstract}
We demonstrate that the extensive class of open Markov quantum systems
describing by the Pauli master equation can be represented in so- called
quasithermodynamic form .Such representation has certain advantages in many
respects for example it allows one to specify precisely the parameter region
in which the relaxation of the system in question to its stationary state
occurs monotonically.With a view to illustrate possible applications of such
representation we consider concrete Markov model that has in our opinion
self-dependent interest namely the explanation of important and well
established by numerous experiments the Yerkes-Dodson law in psychology.
\end{abstract}

\maketitle

\section{Introduction}

The dynamic equations method is the fundamental tool for studying of the
behavior of complex systems in physics,chemistry,population biology and
other sciences. This method can be applied both for the deterministic and
statistical description for the system in question (in the second case the
dynamic equations may be written for the evolution of the probabilities to
find the system in all possible states of its phase space). In the paper 
\cite{1s} we had considered one \ extensive class of dynamical systems so
-called quasithermodynamic systems. We define quasithermodynamic system (QS)
as the system whose behavior can be characterized by two key functions of
its state.By analogy with classical thermodynamics we call these two
functions as the energy and entropy. According to definition these two
functions must satisfy two main conditions (in the first time introduced in
thermodynamics by R. Clausius in 1865, see for example \cite{2s}) that look
as follows:

\textbf{I)} the energy of QS is constant

\textbf{II)} the entropy of QS monotonically increases in time.

Note that for dynamic equations describing various physical and also
nonphysical QS systems the words "energy" and "entropy" should be understand
only in the Pickwick sense as conventional labels for two given functions
satisfying to above mentioned conditions. In the paper \cite{1s} we \
specify the explicit form of dynamic equations for QS whose states are
described by a set of $N$ continuous variables : $x_{1},x_{2}...x_{N}.$ and
examined some important features of their behavior. The main goal of the
present paper to demonstrate that well known Pauli master equation (PME) for
diagonal elements of density matrix of some open quantum Markov system can
be successfully represented in similar quaithermodynamic form. Such
representation brings certain advantages in many respects. In particular as
we prove later in this paper it allows one to specify precisely the
situations when the QS under consideration tends to its stationary (or
equilibrium) state monotonically in time. In addition we consider also one
instructive illustration of such representation relating to psychology that
in our opinion has self-dependent interest.

The paper is organized as follows. In Sect.1 we briefly remind the necessary
facts relating to the theory of QS in particular specify the explicit form
of dynamical equations that provide the realization of the Clausius
conditions \textbf{I),II)}. In Sect.2 \ that is the central part of the
paper we consider the general PME describing the evolution of diagonal
elements of expensive class of open quantum Markov systems and demonstrate
that it can be represented in required quasithermodynamic form. Note that in
the present paper we consider the diagonal elements of density matrix that
is the probabilities $p_{i}$ of finding the system in the state $\left\vert
i\right\rangle $ as basic set of variables.In addition the sum of these
diagonal elements $\sum\limits_{i=1}^{N}p_{i}$ will play the role of energy
in our case. Evidently that in virtue of normalization condition this sum is
conserved and moreover identically equal to unit. The only but nontrivial
problem which remains is the problem of the explicit construction of
corresponding function of entropy that provides the desired equations of
motions for probabilities $p_{i\text{ }}$that is initial PME. Also in this
section we specify the conditions which must be imposed on the Markov system
of interest in order to provide monotonic damping to its stationary state.
In the Sect.3 as some instructive example we study concrete 3 state Markov
model that in our opinion explains one important phenomenon in psychology of
learning namely the Yerkes-Dodson law. Now let us go to the presentation of
concrete results of the paper.

\section{Preliminary information concerning the theory of QS}

In this part we give the brief account relating to the theory of QS, that is
the systems which satisfy the above two Clausius conditions \textbf{I),II)}.
The simplest example of QS is the dynamic system whose state is described by
two continuous variables $\left( x_{1},x_{2}\right) $ and corresponding
equations of the motion may be written in the next form:%
\begin{equation}
\frac{dx_{i}}{dt}=\varepsilon _{ik}\frac{\partial H}{\partial x_{k}}\left\{
S,H\right\} ,  \label{c1}
\end{equation}%
where $H\left( x_{1},x_{2}\right) $ and $S\left( x_{1},x_{2}\right) $ are
two preassigned functions of state, $\varepsilon _{ik}$ is completely
asymmetric tensor of the second rank and $\left\{ f,g\right\} =\varepsilon
_{ik}\frac{\partial f}{\partial x_{i}}\frac{\partial g}{\partial x_{k}}$ is
ordinary Poisson bracket for two functions $f\left( x_{1},x_{2}\right) $ and 
$g\left( x_{1},x_{2}\right) $. It is easy to see directly that equations of
motion Eq. (\ref{c1}) imply the relations: 1) $\frac{dH}{dt}=0$ and 2) $%
\frac{dS}{dt}=\left\{ S,H\right\} ^{2}\geqslant 0$. Hence the functions $H$
and $S$ satisfy to conditions \textbf{I) - II)} and can be considered as
"energy" and "entropy" of corresponding QS. Similarly one can write the
equations of motions for QS with three variables $x_{1},x_{2},x_{3}$ in the
following form:%
\begin{equation}
\frac{dx_{i}}{dt}=\varepsilon _{ikl}\frac{\partial H}{\partial x_{k}}A_{l},
\label{c3am}
\end{equation}%
where the vector $A_{l}=\varepsilon _{lmn}\frac{\partial S}{\partial x_{m}}%
\frac{\partial H}{\partial x_{n}}$ and $\varepsilon _{ikl\text{ \ }}$is
completely antisymmetric tensor of the third rank. Expression Eq. (\ref{c3am}%
) may be rewritten also in the equivalent form:%
\begin{equation}
\frac{dx_{i}}{dt}=\frac{\partial S}{\partial x_{i}}\sum\limits_{k}\left( 
\frac{\partial H}{\partial x_{k}}\right) ^{2}-\frac{\partial H}{\partial
x_{i}}\sum\limits_{k}\left( \frac{\partial H}{\partial x_{k}}\frac{\partial S%
}{\partial x_{k}}\right)  \label{c3b}
\end{equation}%
However it should be noted that expressions Eq. (\ref{c3am}) and Eq. (\ref%
{c3b}) are not the most general form of equations for QS with three
variables. In fact we may add in r.h.s of the Eq. (\ref{c3am}) the
"hamiltonian" term $-r\varepsilon _{ikl}\frac{\partial S}{\partial x_{k}}%
\frac{\partial H}{\partial x_{l}}$ (where $r$ is a multiplier) without any
changing of its quasithermodynamic character. So the general form of QS with
three variables reads as%
\begin{equation}
\frac{dx_{i}}{dt}=\varepsilon _{ikl}\frac{\partial H}{\partial x_{k}}\left(
A_{l}-r\frac{\partial S}{\partial x_{l}}\right) ,  \label{c4}
\end{equation}%
where the vector $A_{l}$ in Eq. (\ref{c4})\ is defined in just the same way
as in Eq. (\ref{c3am}).

The task of description the explicit form of equations of motion for QS with
more than three variables in principle can be solved \ by the same way and
we will turn to it a little later. Now let us draw our attention to the
other important object of present study namely Pauli master equation (PME).
The PME describes the evolution in time the diagonal elements $P_{n}$ of
density matrix of open quantum Markov system (that is the probabilities to
find it in any quantum state $\left\vert n\right\rangle $). This equation
has the next general form \cite{3s}%
\begin{equation}
\frac{dP_{n}}{dt}=\sum\limits_{m}\left( W_{nm}P_{m}-P_{n}W_{mn}\right) ,
\label{c5}
\end{equation}%
where, $W_{nm}$ is a probability (per unit time) of transition from quantum
state $\left\vert m\right\rangle $ to state $\left\vert n\right\rangle $. It
is known that Eq. (\ref{c5}) describes both the relaxation of closed Markov
system to its equilibrium state and the decay of open system to it
nonequilibrium stationary state .In the prominent paper \cite{4s} J.S.
Tomsen proved some important connections existing between symmetry
properties of the coefficients $W_{nm}$ and the character of corresponding
relaxarion process described by master equation Eq. (\ref{c5}). For example
if coefficients $W_{nm\text{ }}$ are symmetric $W_{nm}=W_{mn}$ then all
probabilities $p_{i}^{0}$ in their final stationary state are equal to each
other i.e. the ergodic hypothesis in this case holds. Obviously the symmetry
condition implies the validity of the detailed balance principle: $%
p_{n}^{0}W_{mn}=p_{m}^{0}W_{nm}$ as well. In addition note that the more
weak property of matrix $W_{nm}$ namely its double stochasticity: $%
\sum\limits_{n}W_{mn}=\sum\limits_{n}W_{nm}$ for all indexes $m$ implies
that the Boltzmann-Shennon entropy function $S_{BS}=-\sum\limits_{i}p_{i}\ln
p_{i}$ increases in time (that is $\frac{dS}{dt}\geq 0$). Thus we can
conclude that in symmetric case the PME in fact describes the evolution of
the closed quantum system to its equilibrium state.However in our paper we
are interested in more general case of open nonequilibrium Markov system
when Eq. (\ref{c5}) describes \ its damping to stationary state as well.So
we do not impose in advance any special restrictions on matrix $W_{mn}$. Now
let us turn to our main goal namely to the statement that arbitrary PME can
be represented in the form of appropriate QS.

\section{The representation of the PME in quasithermodynamic form.}

We begin our study with the simplest case of two level open quantum system
that can be described by the PME.Then the PME for the diagonal elements of
its density matrix $\widehat{\rho }$ namely $p_{1}=\rho _{11}$ and $%
p_{2}=\rho _{22}$ \ looks as:%
\begin{eqnarray}
\frac{dp_{1}}{dt} &=&W_{12}p_{2}-p_{1}W_{21},  \notag \\
&&\text{and}  \label{c6} \\
\frac{dp_{2}}{dt} &=&W_{21}p_{1}-p_{2}W_{12}  \notag
\end{eqnarray}%
One can easily verify that the system Eq. (\ref{c6}) may be represented in
required qusithermodynamic form: $\frac{dp_{i}}{dt}=\varepsilon _{ik}\frac{%
\partial H}{\partial p_{k}}\left\{ S,H\right\} $ if we define "energy" $H$ \
as $H=p_{1}+p_{2}$ and "entropy" $\ S$ \ as $S=-\frac{W_{21}p_{1}^{2}}{2}-%
\frac{W_{12}p_{2}^{2}}{2}$.

Note if the symmetry condition $W_{12}=W_{21}$ holds than this "entropy"
function in fact coincides with linear Boltzmann-Shennon entropy that
provides the relaxation of the system to its equilibrium state with $%
p_{1}^{0}=p_{2}^{0}=\frac{1}{2}$. However in general two state Markov system
we have for the final probabilities: $p_{1}^{0}=\frac{W_{12}}{W_{12}+W_{21}}$
and $p_{2}^{0}=\frac{W_{21}}{W_{12}+W_{21}}$ and ergodic hypothesis does not
holds.It is clear that two state case is too simple to shed light on general
case but in the next in complexity three-state case all key elements of
general construction can be guessed. Therefore we consider this case more
detail.For three -level open quantum system the general PME Eq. (\ref{c5})
can be written in the form%
\begin{eqnarray}
\frac{dp_{1}}{dt} &=&-\left( a+b\right) p_{1}+cp_{2}+ep_{3}  \notag \\
\frac{dp_{2}}{dt} &=&ap_{1}-\left( c+d\right) p_{2}+fp_{3}  \label{c7} \\
\frac{dp_{3}}{dt} &=&bp_{1}+dp_{2}-\left( e+f\right) p_{3}  \notag
\end{eqnarray}%
The full coincidance between the PME Eq. (\ref{c5}) an the system of
equations Eq. (\ref{c7}) can be achieved if one introduces the notation: $%
a=W_{21}$, $b=W_{31}$, $c=W_{12}$, $d=W_{32}$, $e=W_{13}$, and $f=W_{23}$.

Note by the way that the general PME for the system with $N$ basic states
obviously has $N\left( N-1\right) $ independent coefficients so in three
state case there are precisely 6 such parameters.Now let us seek a
representation of the PME in required quasithermodynamic form as%
\begin{equation}
\frac{dp_{i}}{dt}=\varepsilon _{ikl}\frac{\partial H}{\partial p_{k}}\left(
A_{l}-r\frac{\partial S}{\partial p_{l}}\right) ,  \label{c8}
\end{equation}%
where all indexes take values $1$, $2$, $3$, the vector $A_{l}=\varepsilon
_{lmn}\frac{\partial S}{\partial p_{m}}\frac{\partial H}{\partial p_{n}}$, $%
H=\sum\limits_{i=1}^{3}p_{i}$ and $r$ is some unknown multiplier. Entropy
function $S\left( p_{1},p_{2},p_{3}\right) $ may be represented as symmetric
quadratic form of basic variables $p_{i}$ that is%
\begin{equation}
S=\frac{Ap_{1}^{2}}{2}+\frac{Bp_{2}^{2}}{2}+\frac{Cp_{3}^{2}}{2}+\alpha
p_{1}p_{2}+\beta p_{1}p_{3}+\gamma p_{2}p_{3}  \label{c9}
\end{equation}%
Note that the transformation: $S\Longrightarrow S+k\left(
p_{1}+p_{2}+p_{3}\right) ^{2}$ does not change equations of motion Eq. (\ref%
{c8}) so without loss of generality we can put the value of $\gamma $ is
equal to zero. Thus in the case of three state Markov system we have 6
unknown coefficients:$A$, $B$, $C$, $\alpha $, $\beta $ and $r$ that
accurately corresponds to 6 parameters $a,$ $b,$ $c,$ $d,$ $e,$ $f$ of
original PME. Now let us determine the explicit connection between PME Eq. (%
\ref{c7}) and its representation in quasithermodynamic form Eq. (\ref{c8}).
Taking into account the above expression for \ the vector $A_{l}$ one can
rewrite Eq. (\ref{c8}) in the next expanded form%
\begin{eqnarray}
\frac{dp_{1}}{dt} &=&2\frac{\partial S}{\partial p_{1}}-\left( 1-r\right) 
\frac{\partial S}{\partial p_{2}}-\left( 1+r\right) \frac{\partial S}{%
\partial p_{3}}  \notag \\
\frac{dp_{2}}{dt} &=&2\frac{\partial S}{\partial p_{2}}-\left( 1-r\right) 
\frac{\partial S}{\partial p_{3}}-\left( 1+r\right) \frac{\partial S}{%
\partial p_{1}}  \label{c10} \\
\frac{dp_{3}}{dt} &=&2\frac{\partial S}{\partial p_{3}}-\left( 1-r\right) 
\frac{\partial S}{\partial p_{1}}-\left( 1+r\right) \frac{\partial S}{%
\partial p_{2}}  \notag
\end{eqnarray}%
Now substituting the expression Eq. (\ref{c9}) for entropy function $S$ in
r.h.s. of Eq. (\ref{c10}) and compare the result with the PME Eq. (\ref{c7})
after a simple algebra we obtain the next relations for unknown coefficients 
$\alpha ,$ $\beta ,$ $B,$ $C,$%
\begin{eqnarray}
\alpha &=&\frac{\left( 1+r\right) c-\left( 1-r\right) d}{3+r^{2}},
\label{c11a} \\
\beta &=&\frac{\left( 1-r\right) e-\left( 1+r\right) d}{3+r^{2}},  \notag \\
B &=&\frac{-2d-\left( 1-r\right) c}{3+r^{2}},C=\frac{-2f-\left( 1+r\right) e%
}{3+r^{2}}.  \notag
\end{eqnarray}%
Besides we have two additional equations that connect coefficients $a$ and $%
b $ from the PME $\left( 7\right) $ with unknown coefficients $A$ and $r:$%
\begin{eqnarray}
a &=&2\alpha -\beta \left( 1-r\right) -\left( 1+r\right) A  \label{c11b} \\
b &=&2\beta -\alpha \left( 1+r\right) -\left( 1-r\right) A.  \notag
\end{eqnarray}%
Substituting expressions Eq. (\ref{c11a}) into Eq. (\ref{c11b}) and equating
two values for coefficient $A$ we obtain the final value for the coefficient 
$r$. If one introduce the notation $\varkappa =\frac{b+e+f}{a+d+e}$ then the
expression for $r$ reads as \ \ $r=\frac{1-\varkappa }{1+\varkappa }.$It is
obvious that if the condition%
\begin{equation}
a+d+e=b+c+f  \label{c12}
\end{equation}%
is valid (that is $\varkappa =1$), the purely "hamiltonian term" $-r\epsilon
_{ikl}\frac{\partial H}{\partial p_{k}}\frac{\partial S}{\partial p_{l}}$ in
quasithermodynamic representation Eq. (\ref{c8}) vanishes.Let us prove now
that condition Eq. (\ref{c12}) implies that relaxation of the three state
open Markov system to its stationary state occurs monotonically. Indeed if
we will search the solutions of linear PME Eq. (\ref{c7}) in standard form
as $p_{i}\left( t\right) =C_{i}e^{\lambda t}$ then after the simple algebra
we obtain the qubic secular equation for three roots of this equation .One
root is precisely equal to zero (since the sum $\sum\limits_{i=1}^{i=3}p_{i}$
is conserved). The other two roots can be obtained from the following
quadratic equation:%
\begin{equation}
\lambda ^{2}+\xi \lambda +\eta \left( a+b+e\right) -\left( e-c\right) \left(
f-a\right) =0  \label{c13m}
\end{equation}%
where, $\xi =a+b+c+d+e+f$, $\eta =c+d+f$. Provided that the determinant of
this equation is lesser than zero two roots of Eq. (\ref{c13m}) will be real
and negative. Thus the necessary and sufficient condition of monotonic
relaxation of open Markov system Eq. (\ref{c7}) to its stationary state may
be written as%
\begin{equation}
\xi ^{2}+4\left( e-c\right) \left( f-a\right) -4\eta \left( a+b=e\right)
\leqslant 0  \label{c14}
\end{equation}%
Let us introduce the notation: $k=e-c,l=f-a,m=b-d$ and $\omega =\left(
a+d+e\right) -\left( b+c+f\right) .$ Then in new notation the condition Eq. (%
\ref{c14}) looks as $\ \omega ^{2}+4\omega \left( l+m\right) +4\left(
l^{2}+m^{2}+lm\right) \leqslant 0$ $\ $or in more convinient form as%
\begin{equation}
\left( \sqrt{3}u+\frac{2}{\sqrt{3}}\omega \right) ^{2}+v^{2}-\frac{\omega
^{2}}{3}\leqslant 0  \label{c15}
\end{equation}%
where $u\equiv l+m$ and $v\equiv l-m$. We see that the boundary of the
region in parameter space of the PME Eq. (\ref{c7}) where the nonmonotonic
relaxation of its solution is possible may be represented by the ellipse: $%
\left( \sqrt{3}u+\frac{2}{\sqrt{3}}\omega \right) ^{2}+v^{2}=\frac{\omega
^{2}}{3}$. Obviously if $\omega =0,$ that is condition $a+d+e=b+c+f$ holds,
the ellipse degenerates into single point and all solutions of Eq. (\ref{c7}%
) monotonically decrese in time. On the other hand if $\omega \neq 0$ there
is a finite region of parameters (the greater the more $\omega $ is) where
nonmonotonic behavior of solutions of Eq. (\ref{c7}) is possible. So the
required result is proved. Now let us discuss in short the case of general
Markov open system that can be described by PME Eq. (\ref{c5}).

.First of all note that above mentioned construction for three state Markov
system can be realized with necessary changes in general case as well.We
propose here only a short outline of complete proof.So let us consider the $%
N $ state Markov system that is described by corresponding PME with $N\left(
N-1\right) $ independent coefficients. We present the QR of the PME for this
system in the next schematic form:%
\begin{equation}
\frac{dp_{i_{1}}}{dt}=\varepsilon _{i_{1}i_{2}...i_{N}}\frac{\partial H}{%
\partial p_{i_{2}}}A_{i_{3}....i_{N}}+\sum\limits_{\alpha =1}^{\frac{\left(
N-1\right) \left( N-2\right) }{2}}r_{\alpha }H_{i_{1}}^{\left( \alpha
\right) }  \label{c16}
\end{equation}%
where $H=\sum\limits_{i=1}^{N}p_{i}$, $A_{i_{3}....i_{N}}=\varepsilon
_{i_{1}i_{2}...i_{N}}\frac{\partial S}{\partial p_{i_{1}}}\frac{\partial H}{%
\partial p_{i_{2}}}$, $S\left( p_{1},..p_{N}\right) $ is symmetric quadratic
form of $N$ variables and $\varepsilon _{i_{1}....i_{N}}$ is completely
antisymmetric tensor of \ N rank .In addition each of the $\frac{\left(
N-1\right) \left( N-2\right) }{2}$ quasihamiltonian terms $\
H_{i_{1}}^{\left( \alpha \right) }$ has the following form:%
\begin{equation}
H_{i_{1}}^{\left( \alpha \right) }=\varepsilon _{i_{1}i_{2}i_{3}....i_{N}}%
\frac{\partial S}{\partial p_{i_{2}}}\frac{\partial H}{\partial p_{i_{3}}}%
R_{i_{4}...i_{N}}^{\left( \alpha \right) }  \label{c17}
\end{equation}%
where every antisymmetric tensor $R_{i_{4}...i_{N}}^{\left( \alpha \right) }$
has $N-3$ rank.The quasithermodynamic representation of Eq. (\ref{c17}) may
be constructed by the next procedure. First of all we examine $N$%
-dimensional vector space representing the states of initial Markov system
in question.Then we consider the subspace consisting from all vectors that
are orthogonal to the vector: $\frac{\partial H}{\partial p_{i}}=\left(
1,1....1\right) .$Obviously this subspace has dimensionality $N-1.$After
that we choose from the basis of this subspace arbitrarily $N-3$ vectors and
form from them by standard way the antisymmetric tensor of $N-3$ rank.Each
of these tensors(with corresponding coefficient $r_{\alpha }$ enters in the
sum in r.h.s. of Eq. (\ref{c16}). It is clear that we can obtain in this way
precisely $C_{N-1}^{N-3}=C_{N-1}^{2}$ distinct antisymmetric terms and
correspondently $C_{N-1}^{2}$ free parameters $r_{\alpha }$. Let us
calculate now the total number of free parameters being at our disposal.The
entropy function as symmetrical quadratic form of $N$ variables gives us $%
\left[ \frac{N\left( N+1\right) }{2}-1\right] $ parameters ( we take into
account that $S$ is defined up to the term $k\left( p_{1}+....p_{N}\right)
^{2}$.Besides due to different choice of quasihamiltonian terms we have $%
C_{N-1}^{2}=\frac{\left( N-1\right) \left( N-2\right) }{2}$ additional
parameters.As the final result we obtain $\frac{N\left( N+1\right) }{2}-1+%
\frac{\left( N-1\left( N-2\right) \right) }{2}=N\left( N-1\right) $ unknown
parameters which enables us uniquiely determine them with the help of
original PME coefficients. QED.

In conclusion of this part note that the existence of entropy function (or
functional in the case of infinite dimensional Markov system) let one the
good possibility to apply powerful variotional methods for study general PME.

\section{Simple Markov quasithermodynamic model may explain the
Yerkes-Dodson Law in psychology.}

In this part as the instructive illustration of forgoing general approach we
consider well known in psychology of learning (see for examle $\left[ 5%
\right] $) the Yerkes-Dodson Law (YDL) which asserts the existence of
optimal level of arousal (or motivation) in learning process and besides the
main feature of this law namely: more complicated the task the lower this
optimal level should be.We propose here the simple Markov model that in our
opinion explains The YDL qualitatively and in first approximation
quantitatively as well.In order to explain the YDL we assume as correct the
hypothesis of functional equavalence between perception and other human
cognitive processes including the learning process $\left[ 6\right] $.Remind
that sensory information processing in brain occurs in two subsequent
steps.In an initial stage (segmentation) certain groups of similar features
of perceived object form so-called clusters of perception and in the second
stage (binding) these separated clusters are integrated into complete
perceptual image.Analogously we assume that the states of training
individual during the learning process can be characterized by the following
way .There are three basic states: untrained state $\left\vert
1\right\rangle $, poorly trained state $\left\vert 2\right\rangle $,and well
- trained state $\left\vert 3\right\rangle $.Also we suppose that for
relevant description of YDL in learning it is enough to take into account
two successive stage of learning namely a) the primary learning i.e. the
transition $\left\vert 1\right\rangle \Longrightarrow $ $\left\vert
2\right\rangle ,$and b) the secondary or high learning i.e. the transition $%
\left\vert 2\right\rangle $ $\Longrightarrow \ \left\vert 3\right\rangle ,$%
and in addition two destructive transitions that impeding to successful
learning c) partial loss of the habit in view of exsessive agitation or
various external noise i.e. the transition $\left\vert 3\right\rangle
\Longrightarrow \left\vert 2\right\rangle $ and the inevitable forgetting of
the habit in view of (for example) long absence from practice i.e.the
transition $\left\vert 3\right\rangle $ $\Longrightarrow \left\vert
1\right\rangle .$Now let us formulate the Markov model based on PME that
takes into account all above listed reasons.We believe that relevant
equations of this model can be written in the following way

\begin{eqnarray}
\frac{d\rho _{1}}{dt} &=&-a\rho _{1}+e\rho _{3},  \notag \\
\frac{d\rho _{2}}{dt} &=&a\rho _{1}-d\rho _{2}+f\rho _{3}  \label{c18} \\
\frac{d\rho _{3}}{dt} &=&d\rho _{2}-\left( e+f\right) \rho _{3}  \notag
\end{eqnarray}%
where the coefficients $a$, $d$, $e$, $f$ descrbe the probabilities (per
unit time) of above mentioned transitions.We consider $\rho _{i}$ $\left(
i=1,2,3\right) $ as the probabilities to find the individual in
corresponding state of learning.Comparing the Eq. (\ref{c1}) with general
three state PME Eq. (\ref{c7}) we see that the model proposed Eq. (\ref{c18}%
) corresponds to its partial case when coefficients $b=c=0$. It is easy to
see that the stationary solution of Eq. (\ref{c18}) has the form:%
\begin{eqnarray}
\rho _{1}^{0} &=&\frac{de}{de+a\left( d+e+f\right) },  \notag \\
\rho _{2}^{0} &=&\frac{a\left( e+f\right) }{de+a\left( d+e+f\right) },
\label{c19m} \\
\rho _{3}^{0} &=&\frac{ad}{de+a\left( d+e+f\right) }  \notag
\end{eqnarray}%
Up to this point we did not take into account the influence of arousal (or
motivation) on learning process. Now let us do it.On the grounds of simple
psychological reasons we believe that increase of arousal promotes only the
transitions $\left\vert 1\right\rangle $ $\Longrightarrow \left\vert
2\right\rangle $ and $\left\vert 3\right\rangle $ $\Longrightarrow
\left\vert 2\right\rangle $ and has minor effect on transitions $\left\vert
3\right\rangle $ $\ \Longrightarrow \left\vert 1\right\rangle $ and $\
\left\vert 2\right\rangle $ $\Longrightarrow \left\vert 3\right\rangle $. If
we denote the arousal level of training individual (which can be measured by
relevant psychological methods as $k$ ), then our assumptions can\ be
explicily expressed in the form of next two relations: $a=a_{1}k$ and $%
f=f_{1}k$ $.$ Now we believe that coefficients $a_{1},f_{1},d,e$ do not
depend on arousal. Finally the probability to find the individual in
stationary well-trained state can be obtained from Eq. (\ref{c19m}) and
looks as%
\begin{equation}
\rho _{3}^{0}=\frac{a_{1}dk}{de+a_{1}\left( d+e\right) k+a_{1}f_{1}k^{2}}
\label{c20}
\end{equation}%
The maximum of expression Eq. (\ref{c18}) is reached when the arousal level
is equal to%
\begin{equation}
k_{ext}^{2}=\frac{de}{a_{1}f_{1}}.  \label{c21}
\end{equation}%
It is also worth noting that in the cases when the the errors in learning
can result in grave consequences (for example in such professions as surgeon
or pilot) it is highly desirable that the learning process would be
consistent .To this end the instructor during the learning process must try
to provide the fulfilment of two conditions 1) providing optimal level of
motivation that is $k_{opt}=\sqrt{\frac{de}{a_{1}f_{1}}}$ and 2) that
warrants serious failures in training $:\frac{d+e}{\sqrt{de}}=\frac{%
f_{1}-a_{1}}{\sqrt{f_{1}a_{1}}}$.The second of these conditions in fact
entirely coincides with condition Eq. (\ref{c12}).

In conclusion of this part we want to emphasize that all results obtained in
this simplified model of learning process\ undoubtely need in careful
experimental checking and verification.

\end{document}